\documentclass[aps,pra,showpacs,english,11pt,a4,nofootinbib,notitlepage,onecolumn,superscriptaddress]{revtex4-1}
\usepackage{stmaryrd}
\usepackage{amsfonts}
\usepackage{mathrsfs}
\usepackage{amsmath}
\usepackage{amscd}
\usepackage{graphicx}
\usepackage{booktabs}
\usepackage{leftidx}
\usepackage{bm}
\usepackage{braket}
\usepackage{color}
\usepackage{times}

\usepackage{hyperref}
\hypersetup{  colorlinks=true, linkcolor=blue, citecolor=red, urlcolor=blue  }

\begin{document}

\title{Experimental Study of \textsc{Forrelation} in Nuclear Spins}

\author{Hang Li}
\thanks{These authors contributed equally to this work.}
\affiliation{State Key Laboratory of Low-Dimensional Quantum Physics and Department of Physics, Tsinghua University, Beijing 100084, China}
\affiliation{Collaborative Innovation Centre of Quantum Matter, Beijing 100084, China}
\author{Xun Gao}
\thanks{These authors contributed equally to this work.}
\affiliation{Centre for Quantum Information, Institute for Interdisciplinary Information Sciences, Tsinghua University, Beijing 100084, China}
\author{Tao Xin}
\affiliation{State Key Laboratory of Low-Dimensional Quantum Physics and Department of Physics, Tsinghua University, Beijing 100084, China}
\affiliation{Collaborative Innovation Centre of Quantum Matter, Beijing 100084, China}
\author{Man-Hong Yung}
\email{yung@sustc.edu.cn}
\affiliation{Department of Physics, South University of Science and Technology of China, Shenzhen, Guangdong 518055, China}
\affiliation{Centre for Quantum Information, Institute for Interdisciplinary Information Sciences, Tsinghua University, Beijing 100084, China}
\author{Guilu Long}
\email{gllong@mail.tsinghua.edu.cn}
\affiliation{State Key Laboratory of Low-Dimensional Quantum Physics and Department of Physics, Tsinghua University, Beijing 100084, China}
\affiliation{Collaborative Innovation Centre of Quantum Matter, Beijing 100084, China}

\date{\today}

\begin{abstract}
Correlation functions are often employed to quantify the relationships among interdependent variables or sets of data. Recently, a new class of correlation functions, called \textsc{Forrelation}, has been introduced by Aaronson and Ambainis for studying the query complexity of quantum devices. It was found that there exists a quantum query algorithm solving 2-fold \textsc{Forrelation} problems with an exponential quantum speedup over all possible classical means, which represents essentially the largest possible separation between quantum and classical query complexities. Here we report an experimental study probing the 2-fold and 3-fold \textsc{Forrelations} encoded in nuclear spins. The major experimental challenge is to control the spin fluctuation to within a threshold value, which is achieved by developing a set of optimized GRAPE pulse sequences. Overall, our small-scale implementation indicates that the quantum query algorithm is capable of determine the values of \textsc{Forrelations} within an acceptable accuracy required for demonstrating quantum supremacy, given the current technology and in the presence of experimental noise.
\end{abstract}


\maketitle

\section{Introduction}

With the ability of creating exponential number of superposition of states, quantum computation provides an unprecedented computational power over classical computation. For example, Shor's factoring algorithm~\cite{shor1994algorithms}, the Harrow-Hassidim-Lloyd (HHL) algorithm~\cite{harrow2009quantum}, and other progresses in quantum simulation~\cite{Lloyd1996,Yung2012c,Kassal2011} provide strong evidences that quantum computation can gain exponential speed-up in practical problems. Apart from computational decision problems, quantum devices can be exploited for other classically-intractable computational tasks, including sampling distributions of some quantum systems~\cite{aaronson2011computational,bremner2011classical,bremner2016achieving,gao2016quantum,boixo2016characterizing}. As a result, one may expect to gain ``quantum supremacy"~\cite{preskill2012quantum} in relatively-simple quantum devices in the near future.

Although these results are promising, complete and rigorous proofs for supporting the claims of gaining quantum supremacy are still unavailable. Recall that for the case of Shor's algorithm, we have not excluded the possibility for the existence of a polynomial-time classical algorithm for the factoring problem.  For the HHL algorithm, which is \textsf{BQP}-complete, it remains to determine if quantum computation is indeed more powerful than classical computation, or technically, if it is true that $\textsf{BQP} \supset \textsf{BPP}$. Furthermore, the success of the sampling algorithms is founded on several conjectures in the theory of classical computational complexity. Even though boson-sampling devices are capable of creating an exponential number of superposition of quantum states, the transition amplitudes can still be estimated by classical devices within additive errors~\cite{Yung2016}.

On the other hand, query complexity, which counts the number of queries of black-box functions (i.e., without the knowledge of the internal structure), provides further evidences supporting quantum speed-up over the classical counterparts. For example, Grover's search  algorithm~\cite{grover1996fast}, the Deutsch-Jozsa algorithm~\cite{deutsch1992rapid} and Simon's algorithm~\cite{simon1997power} are all characterized in the context of query complexity.

Recently, Aaronson and Ambainis \cite{aaronson2015forrelation} introduced a new concept in  query complexity, called \textsc{Forrelation}, which characterizes the multi-fold correlations among different boolean functions. It was found that a quantum computer is capable of solving 2-fold \textsc{Forrelation} problems within a constant $O(1)$ number of queries. However, classical computers require an exponential number of queries. The difference between the query complexity between quantum and classical methods is shown to be a maximally-achievable separation with quantum methods (see also Refs.~\cite{aaronson2010bqp,de2002sharp,chakraborty2010new}). Furthermore, multiple-fold \textsc{Forrelation} problems are as hard as quantum computation~\cite{aaronson2015forrelation}, i.e., \textsf{BQP}-complete.

\begin{figure}
\begin{center}
\includegraphics[width=10cm,angle=0]{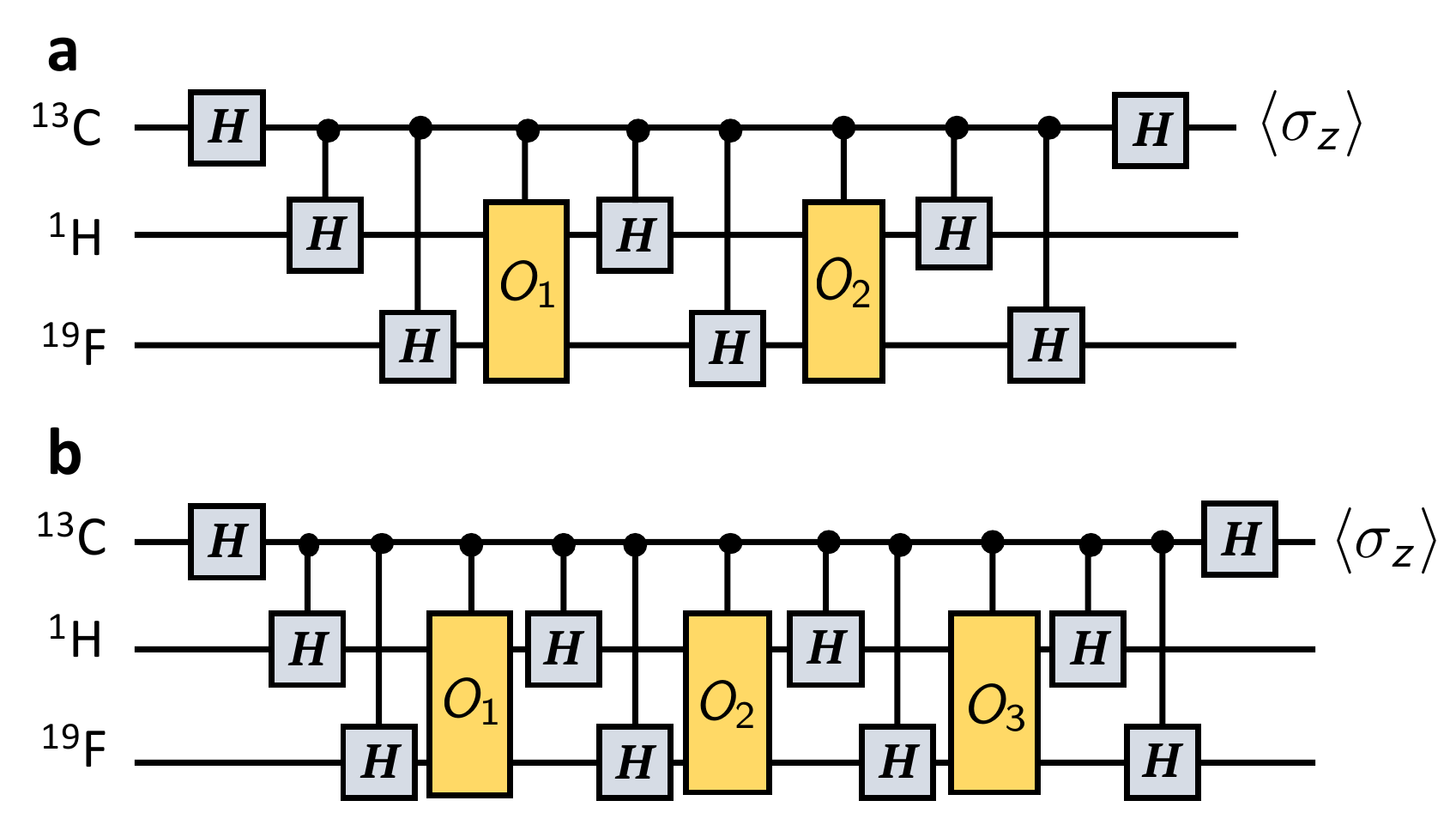}
\caption{Quantum circuit for probing (a) 2-fold and (b) 3-fold \textsc{Forrelation} problems. The system is prepared at state $\ket{000}$. $O_1 \equiv O_{f_{1}}$, $O_2 \equiv O_{f_{2}}$, and $O_3 \equiv O_{f_{3}}$ are query operators that map states $\ket{x}$ to $f_{1}(x)\ket{x}$, $f_{2}(x)\ket{x}$, and $f_{3}(x)\ket{x}$ respectively, where$ f_{1}(x), f_{2}(x), f_{3}(x) \in \{1, -1\}$. }\label{f1}
\end{center}
\end{figure}

Here we report the first experimental study of the 2-fold and 3-fold \textsc{Forrelations} in a system of nuclear spins, where the NMR quantum circuit of 2-fold \textsc{Forrelation} involves only 2 queries of the black box functions, but classically, it takes a total of 8 queries for an exact result. Similarly, 3 queries are needed for the NMR implementation of 3-fold \textsc{Forrelation}, while 12 queries are needed classically if memory is given for the black-box functions; otherwise it can go up to 192 classical queries. 

However, we note that the measurement results come directly form the NMR signals, but a standard implementation of the quantum circuit involves probabilistic measurement outcomes. Furthermore, similar to other experimental demonstrations of Deutsch-Jozsa algorithms~\cite{chuang1998experimental,jones1998implementation}, the applied NMR pulse sequences depends on the knowledge of the functions, which are not strictly ``black boxes". Therefore, the current experimental results cannot be taken as a direct proof for demonstrating quantum supremacy, which is relevant only in the large-$N$ limit. 

The purpose of the experiment is to investigate whether a small-size prototype experiment can produce \textsc{Forrelation} within the accuracy required for demonstrating the quantum advantages (above the threshold $3/5$ or below the threshold $1/100$), given the current technology and in the presence of experimental noise. In particular, our experimental fluctuation for the spin measurement has to be controlled within $1 \%$. These experimental results allow us to identify the places one can improve for scaling up the size of the experiment in future.

\section{\textsc{Forrelation}}

Given $k$ Boolean functions, $f_1 \equiv f_1(x_1),\cdots, f_k \equiv f_k(x_k)$, each with $n$ variables, i.e., $ x_j \in {\left\{ {0,1} \right\}^n} \to \left\{ { - 1,1} \right\}$, the $k$-fold \textsc{Forrelation},  ${\Phi _k} \equiv \Phi_{f_1,f_2,...,f_k}$, of these functions is defined as follows,
\begin{equation}
{\Phi}_k \equiv \sum\limits_{{x_1},{x_2},...} {\frac{{{e^{i\phi ({x_1},{x_2},...)}}}}{{{2^{(k + 1)n/2}}}}{f_1}({x_1}){f_2}({x_2}) \cdots {f_k}({x_k})} \ ,
\end{equation}
where ${e^{i\phi ({x_1},{x_2},...)}} \equiv {( - 1)^{{x_1} \cdot {x_2}}}{( - 1)^{{x_2} \cdot {x_3}}} \cdots {( - 1)^{{x_{k - 1}} \cdot {x_k}}}$, and $x\cdot y$ indicates the bitwise inner product between the $n$-dimensional binary vectors $x$ and $y$. The total number of possible assignment is $N=2^n$. Essentially, ${2}$-fold \textsc{Forrelation} is simply the inner product between a boolean function and Fourier transformation of another boolean function, i.e.,
\begin{equation}
{\Phi _{f,g}} \equiv \frac{1}{{{2^{3n/2}}}}\sum\limits_{x,y \in {{\{ 0,1\} }^n}} {{{( - 1)}^{x \cdot y}}f(x)g(y)} \ .
\end{equation}

Importantly, an exact determination of 2-fold \textsc{Forrelation} ${\Phi _{f,g}}$ is a computationally-hard problem for classical devices, which can be justified by the following challenge~\cite{aaronson2015forrelation}: given a pair of Boolean functions $f$ and $g$, suppose it is known that either (i) $|\Phi_{f,g}|\le1/100$ or (ii) $\Phi_{f,g}\ge 3/5$ is true, all classical methods require an exponential number $\Omega(\sqrt{N}/\log N)$ of queries to the black-box functions, but the quantum computers can finish the task with a constant number of queries. The separation between the quantum and classical query complexity is (almost) largest possible one can achieve~\cite{aaronson2015forrelation}.

\begin{figure}
\begin{center}
\includegraphics[width=\columnwidth, angle=0]{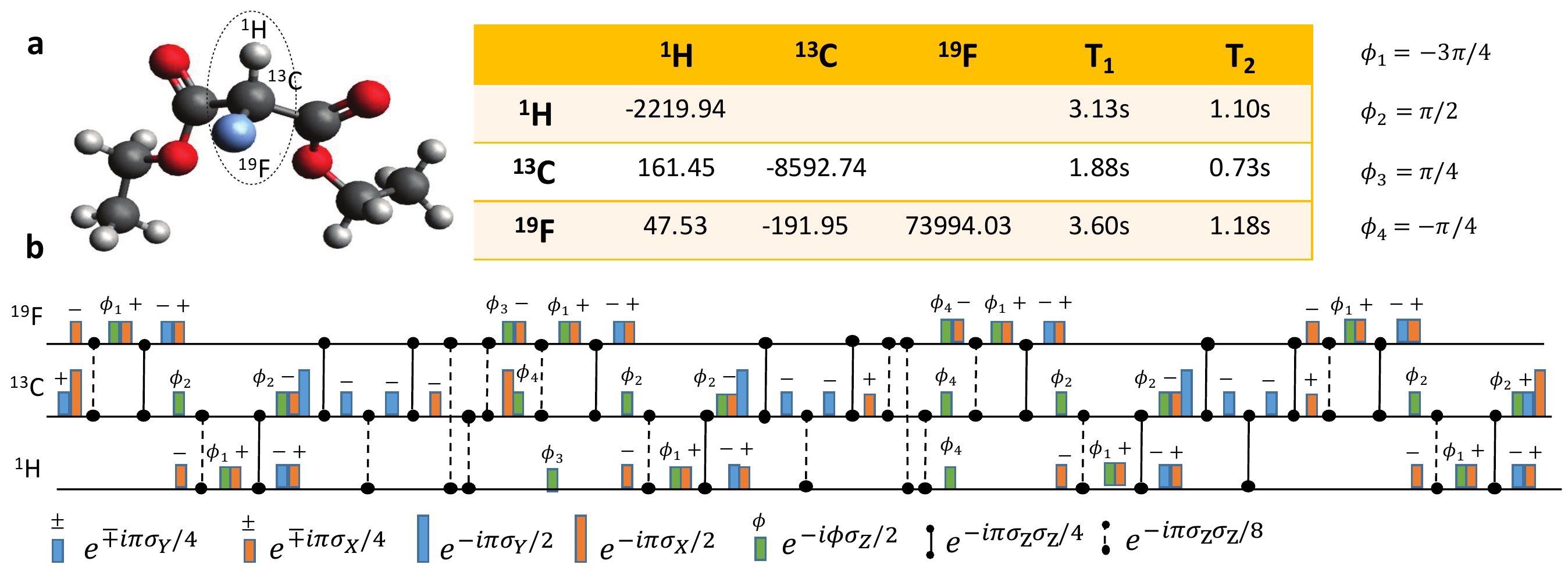}
\caption{(a) Molecular structure and Hamiltonian parameters of Diethyl-fluoromalonate. The chemical shifts and scalar coupling constants of the molecule are on and below the diagonal (in Hz) in the table, respectively. (b) An example of pulse sequences for solving the 3-fold Forrelation problem. $^{13}$C acts as the probe qubit while $^{1}$H and $^{19}$F are the work qubits. The circuit is comprised of 52 $\pi/2$ ($\pi$) hard pulses and 31 free evolution periods under spin-spin J-couplings in total (refocusing pulses are omitted for clarity). }\label{f2}
\end{center}
\end{figure}

Quantum circuits for solving 2-fold and 3-fold \textsc{Forrelation} problems~\cite{aaronson2015forrelation} are shown in Fig.~\ref{f1}. For 2-fold \textsc{Forrelation} problems, there are 2 query operators $O_{f_{1}}$ and $O_{f_{2}}$, which map each input basis state $\ket{x}$ to $f_{1}(x)\ket{x}$ and $f_{2}(x)\ket{x}$ respectively, i.e., ${O_{{f_k}}}\left| x \right\rangle  = {f_k}\left( x \right)\left| x \right\rangle$.

\section{Experimental background} 

Nuclear magnetic resonance (NMR) is a reliable technology for studying small-to-medium size quantum information experiments~ \cite{peng2015zeros,lu2015clifford}, and quantum simulation~\cite{feng2013tunnel,peng2005phase,auccaise2011discord,zhang2012magnet}. Motivated by the needs of studying quantum information, many sophisticated techniques of controlling nuclear spins have been developed. 

Here all the experiments are carried out at room temperature (295 K) on a Bruker Avance III 400 MHz spectrometer and the $^{13}$C labelled Diethyl-fluoromalonate dissolved in $d6$ acetone is used as a 3-qubit NMR quantum information processor. The structure and Hamiltonian parameters of Diethyl-fluoromalonate are shown in Fig.~\ref{f2} (a) where $^{13}$C, $^1$H and $^{19}$F nuclear spins respectively act as an ancillary qubit and two work qubits. Moreover, The internal Hamiltonian of the system is given by

\begin{equation}
H_{int}=\sum\limits_{i=1}^3{\pi\nu_i\sigma^{i}_{z}}+\sum\limits_{j<k,=1}^3{\frac{\pi}{2}J_{jk}\sigma^j_z\sigma^k_z} \ ,
\end{equation}

The whole experimental procedure consists of three parts: (i) state initialization, (ii) realization of the quantum algorithm for solving 2 (or 3)-fold Forrelation problem, and (iii) readout of the expectation value of $\sigma_z^1$ of the ancillary qubit $^{13}C$, which is equal to the \textsc{Forrelation}, i.e.,
\begin{equation}
\left\langle {\sigma _z^1} \right\rangle =  \Phi_k \ ,
\end{equation}
for any $k \ge 2$. We note that for the NMR quantum computing, the whole system, starting from the thermal equilibrium state, is in fact initialized at the pseudo-pure state (PPS) \cite{gershenfeld1997qc,knill2000algorithmic} $\rho_{000}=(1-\varepsilon)I/8+\varepsilon\ket{000}$, using the spatial average technique \cite{cory1997spatial}. To check the success of preparing the PPS, a full quantum state tomography (QST) \cite{lee2002nmr} is carried out. The fidelity between the density matrix prepared in experiment ($\rho_{\rm exp}$) and the target one in theory ($\rho_{\rm th}$) is given by the following expression, $F(\rho_{\rm exp},\rho_{\rm th})\equiv {\rm tr}(\rho_{\rm exp}\rho_{\rm th})/\sqrt{{\rm tr}(\rho_{\rm exp}^2){\rm tr}(\rho_{\rm th}^2)}$. 

A spectrum of the PPS observed on $^{13}$C is shown in Fig.~\ref{f5}(a). The real parts of the initial state are shown in the appendix part. Overall, the initial state can be well prepared in our setup; the fidelity can reach up to 96.9\%.

\begin{figure*}
\begin{center}
\includegraphics[width=\columnwidth, angle=0]{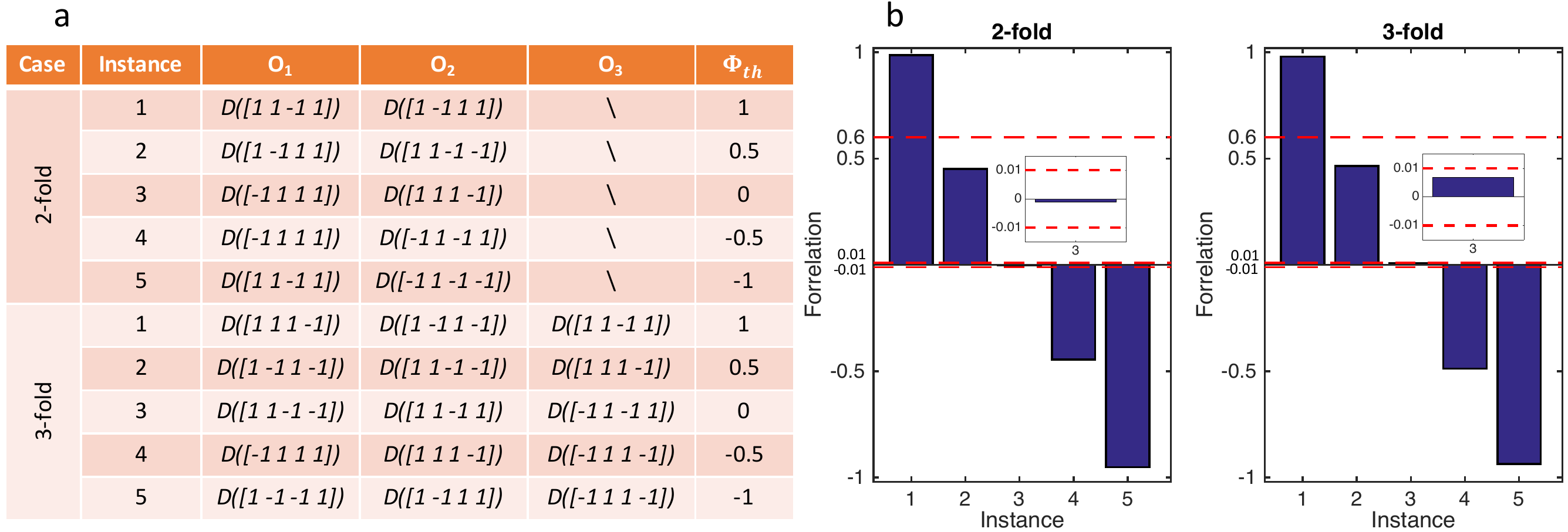}
\caption{(a) Table of 10 selected experimental instances for 2-fold and 3-fold Forrelation problems and their respective target Forrelation $\Phi_{th}$, where {\it D([a b c d])} indicates a $4\times4$ diagonal matrix with diagonal elements {\it a}, {\it b}, {\it c} and {\it d}.  (b) The experimental results of Forrelation $\Phi$. The dashed lines are the criterion vaules for the Forrelation decision problem. The inset is zooming in on the 3rd instance in both plot. Only instance 1 and 3 satisfy $|\Phi|\le1/100$ or $\Phi\ge 3/5$ in both 2-fold and 3-fold case.}\label{f3}
\end{center}
\end{figure*}

\section{Experimental details}
 
To solve the $k$-fold \textsc{Forrelation} problem, a quantum circuit is designed to obtain Forrelation $\Phi_{k} \equiv \Phi_{f_1,...,f_k}$ by measuring the probability of the ancillary qubit in state $\ket{0}$. Here we focus on the experimental results of 2-fold and 3-fold \textsc{Forrelation} in nuclear spins. There are totally five possible values for the \textsc{Forrelation} $\Phi_{f_1,...,f_k}$ in both cases, namely, $\{ 1, 0.5, 0, -0.5,-1 \}$. For each theoretical value of $\Phi_{f_1,...,f_k}$, we associate it with a set of functions listed in the table of Fig.~\ref{f3}(a). There, the operators $O_1$, $O_2$ and $O_3$ are $4\times4$ diagonal matrices in the computational basis, i.e., ${D([a, b, c, d])}$, where $a, b, c, d \in\{1, -1\}$.
 
\begin{figure}[b]
\begin{center}
\includegraphics[width=12cm,angle=0]{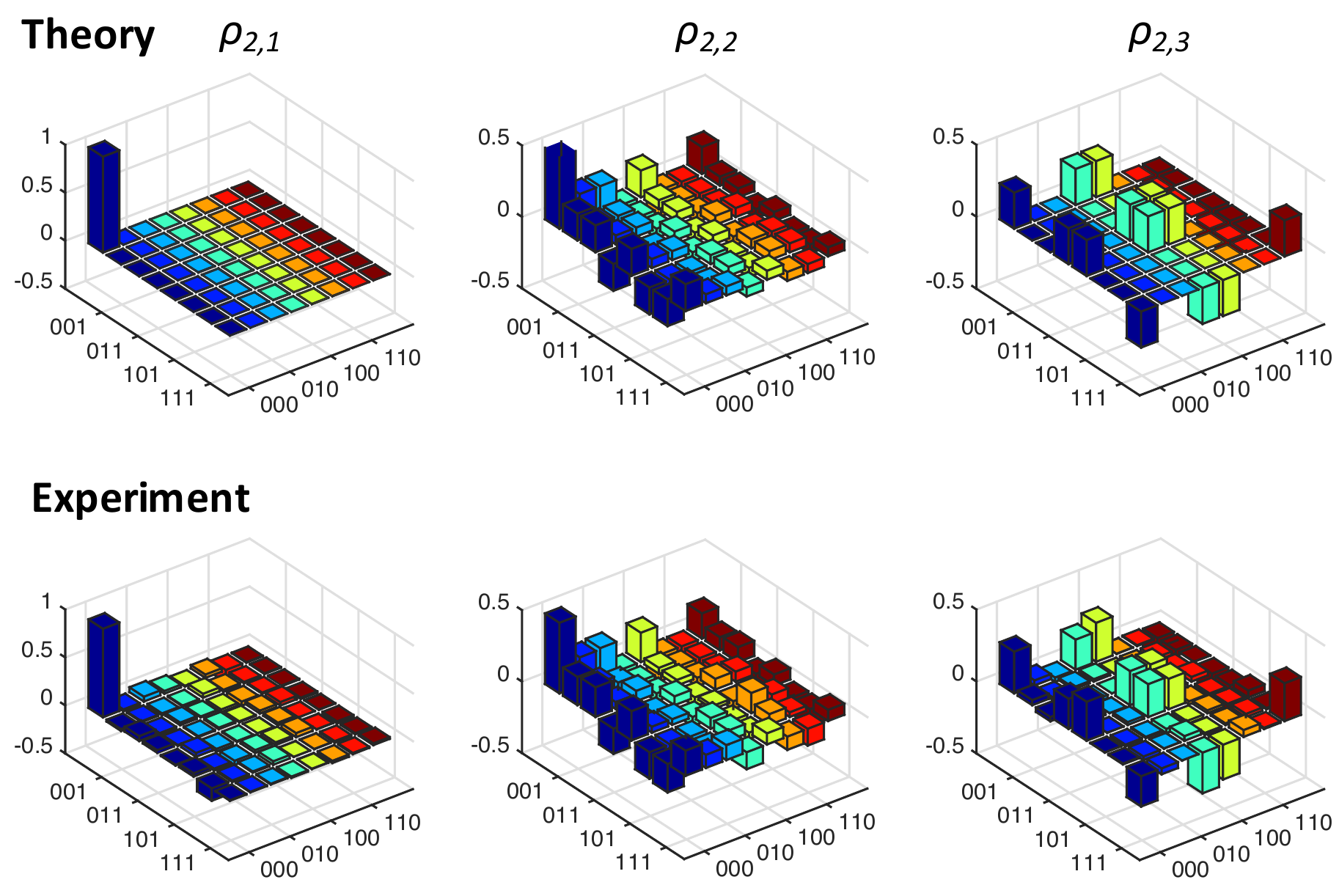}
\caption{Real parts of theoretical and experimental density matrices of $\rho_{2,1}$, $\rho_{2,2}$ and $\rho_{2,3}$. The imaginary parts of experimental density matrices are alomost zero and accordant to the theoretical, which are not presented here due to limited space. In 2-fold case, instance 1 and 3 satisfy $|\Phi|\le1/100$ or $\Phi\ge 3/5$, while instance 2 doesn't.}\label{f4}
\end{center}
\end{figure}

The quantum algorithm for solving the $k$-fold \textsc{Forrelation} problem can be decomposed into several elemental $\pi/2$ (and $\pi$) hard pulses and evolutions under spin-spin J-couplings of the internal Hamiltonian in experiment \cite{vandersypen2005nmr,ryan2008compiler,nielsen2010quantum}, and the whole pulse sequences can be compiled with phase tracking and numerical optimization of the refocusing scheme. For example, the pulse sequences for the instance 4 in the 3-fold case are shown in Fig.~\ref{f2}(b). There are at least 52 $\pi/2$ ($\pi$) hard pulses (without adding the refocusing pulses) and 31 free evolution periods in total are required. In experiment, we utilized the gradient ascent pulse engineering (GRAPE) method \cite{khaneja2005grape} to pack the whole algorithm for each instance into one shaped pulse, with the length of each pulse being 15 ms and the number of segments being 5000. All the shaped pulses are calculated with their fidelities reaching $99.5\%$ and are guaranteed to be robust to the inhomogeneity of radio-frequency pulses.

\section{Experimental results} 

Finally, an observation of the final state on the probe qubit $\langle\sigma_z\rangle$ is conducted in each run of the experiment to get the probabilities of the probe qubit in the $\ket{0}$ state. Fig.~\ref{f3} shows the NMR spectra of the probe qubit $^{13}$C, where (a) is the spectrum of $^{13}$C after a readout pulse when the system is initialized in PPS taken as a calibration. (b), (c) and (d) are the $^{13}$C spectra of the final state after conducting quantum algorithms of the selected transformation operators instance 1, 2 and 3 in 2-fold case, respectively, while (e) and (f) show the spectra after conducting quantum algorithms of the selected transformation operators group 4 and 5 in 3-fold case, respectively. 

The experimental results of the 5 selected instances in 2-fold (3-fold) case are respectively 0.9867, 0.4509, -0.0011, -0.4454 and -0.9516 (0.9791, 0.4659, -0.0068, -0.4871 and -0.9355), as shown in Fig.~\ref{f3}(b). From the results, we can distinguish the 10 selected instances which set they belong. i.e., instance 1 is in the case of  $\Phi>3/5$, instance 3 is in the case of $|\Phi|<1/100$ with bounded probability of error in both 2-fold and 3-fold case; therefore our experimental results indicate that 2-fold and 3-fold \textsc{Forrelation} problems can be solved by making only $1$ quantum queries to each $f_1,f_2$ and $f_3$.

Furthermore, we performed a full quantum state tomography (QST) on the final states.  We conducted QST on instance 1, 2, and 3 in 2-fold case, and instance 4 and 5 in the 3-fold case. To describe the density matrices of the final states, we label $\rho_{k,n}$ for instance $n$ in $k$-fold case. The real parts of the density matrices for the final states of $\rho_{2,1},\rho_{2,2}$ and $\rho_{2,3}$ are presented in Fig.~\ref{f4} (the imaginary parts are very close to zero). A full QST of $\rho_{3,4}$ and $\rho_{3,5}$ are shown in the appendix part. Since the fidelities of all the shaped pulses generated by GRAPE are almost $99.5\%$, the experimental final density matrices are indeed very close to the theoretical ones. The five selected experimental fidelities are $95.27\%$, $96.26\%$, $94.69\%$, $94.73\%$ and $94.64\%$, respectively, indicating a very well implementation of the quantum algorithm in experiment.

\section{Conclusions} 
In summary, we tested a quantum implementation for solving the $k$-fold \textsc{Forrelation} problem~\cite{aaronson2015forrelation} in a prototype experiment. The Forrelation $\Phi_{f_1,...,f_k}$ among a set of ``black-box" functions are obtained by the spin polarization $\langle\sigma_z\rangle$ of an ancillary qubit. The goal of the experiment is to determine if $|\Phi_{f,g}|\le1/100$ or $\Phi_{f,g}\ge 3/5$. In our experiments, 5 selected instances of both the {2}-fold and {3}-fold \textsc{Forrelation} problems are solved on a three-qubit NMR quantum information processor. The experimental results successfully identify that instance 1 and 3 are in the case of $|\Phi_{f,g}|\le1/100$ or $\Phi_{f,g}\ge 3/5$ in both 2-fold and 3-fold case. The quality in the preparation of the PPS and the implementation of the quantum algorithm are benchmarked by a full quantum state tomography for both the initial and the final states. Besides, all the shaped pulses are designed to be robust to the inhomogeneity of the radio-frequency pulses. The main source of errors are caused by the imperfection of GRAPE pulses and the instrumental-related imperfection of the shaped pulse. The total length of each shaped GRAPE pulse is only 15 ms, which is much shorter than the relaxation time of our system.  To our knowledge, this is the first implementation of solving \textsc{Forrelation} problem reported in the literature, and the experimental method can be extended for more complex version of the multiple-fold \textsc{Forrelation} in $2^n$-dimensional space, and be implemented in other platforms such as superconducting devices and trapped ions.

\begin{acknowledgments}
The authors would like to thank D. Lu, S. Hou and G. Feng for helpful discussions, we also thank IQC, University of Waterloo, for providing the software package for NMR experiment simulation. This work is supported by the National Natural Science Foundation of China under Grant No. 11175094, 91221205, and 11405093, and the National Basic Research Program of China under Grants No. 2015CB921002.
\end{acknowledgments}

\section*{Appendix: Experimental spectra and other state reconstructions}

\begin{figure}
\begin{center}
\includegraphics[width=\columnwidth, angle=0]{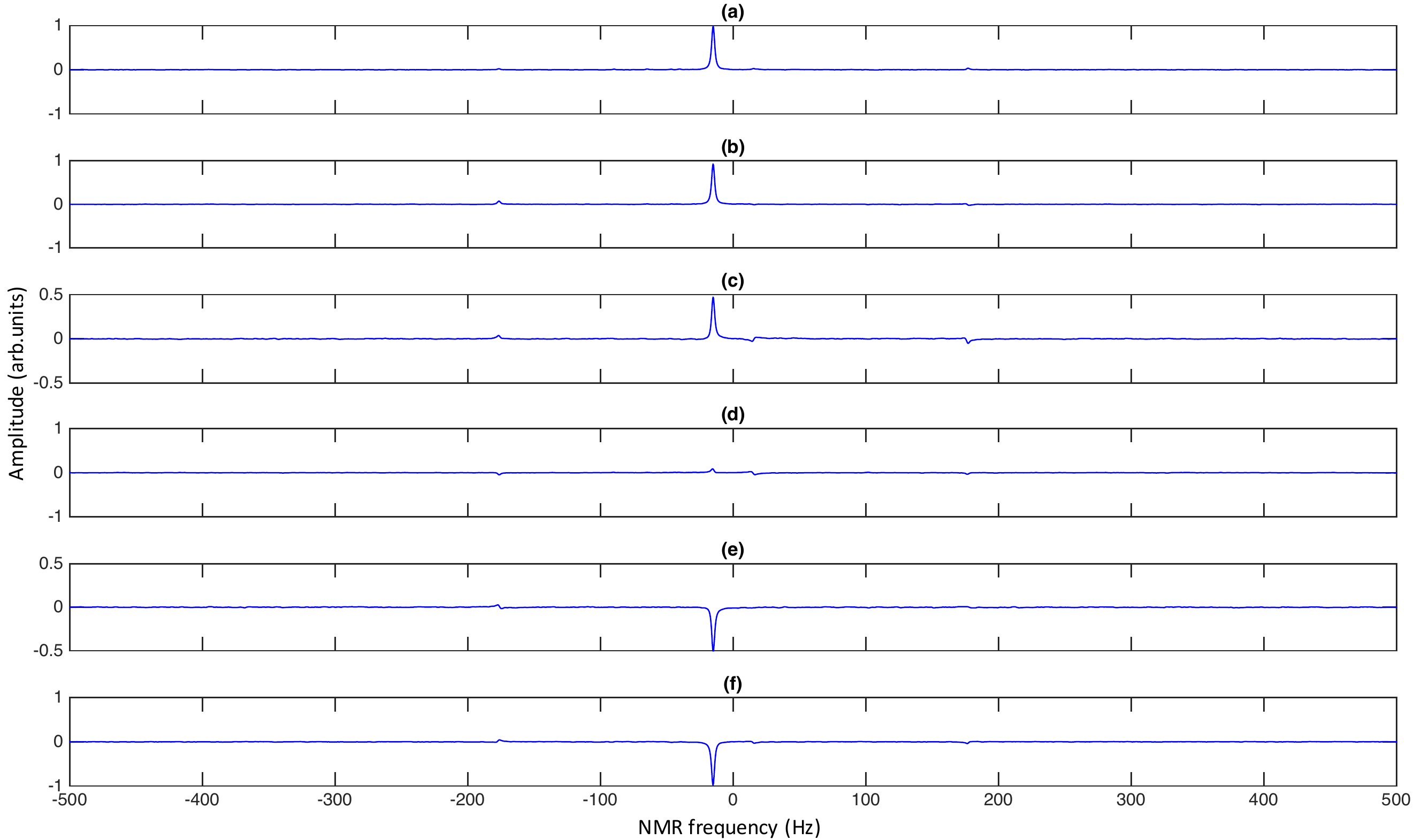}
\caption{NMR spectra of the probe qubit $^{13}$C after a readout pulse. (a) Spectrum of the PPS. (b), (c) and (d) Spectra of experimental instance 1,  2 and 3 in 2-fold case, respectively. (d) and (e) Spectra of experimental instance 4 and 5 in 3-fold case, respectively.}\label{f5}
\end{center}
\end{figure}

All the experimental data are gathered from the probe qubit $^{13}C$, measuring the operator $\langle\sigma_z\rangle$ on $^{13}C$. A combination of 6 different $^{13}C$ NMR spectra is presented in Fig.~\ref{f5} to give a direct and clear comparison, including a PPS spectrum and 5 NMR spectra after implementing a \textsc{Forrelation} circuit, which correspond to 5 different \textsc{Forrelation} value 1, 0.5, 0, -0.5 and -1.

\begin{figure}
\begin{center}
\includegraphics[width=12cm, angle=0]{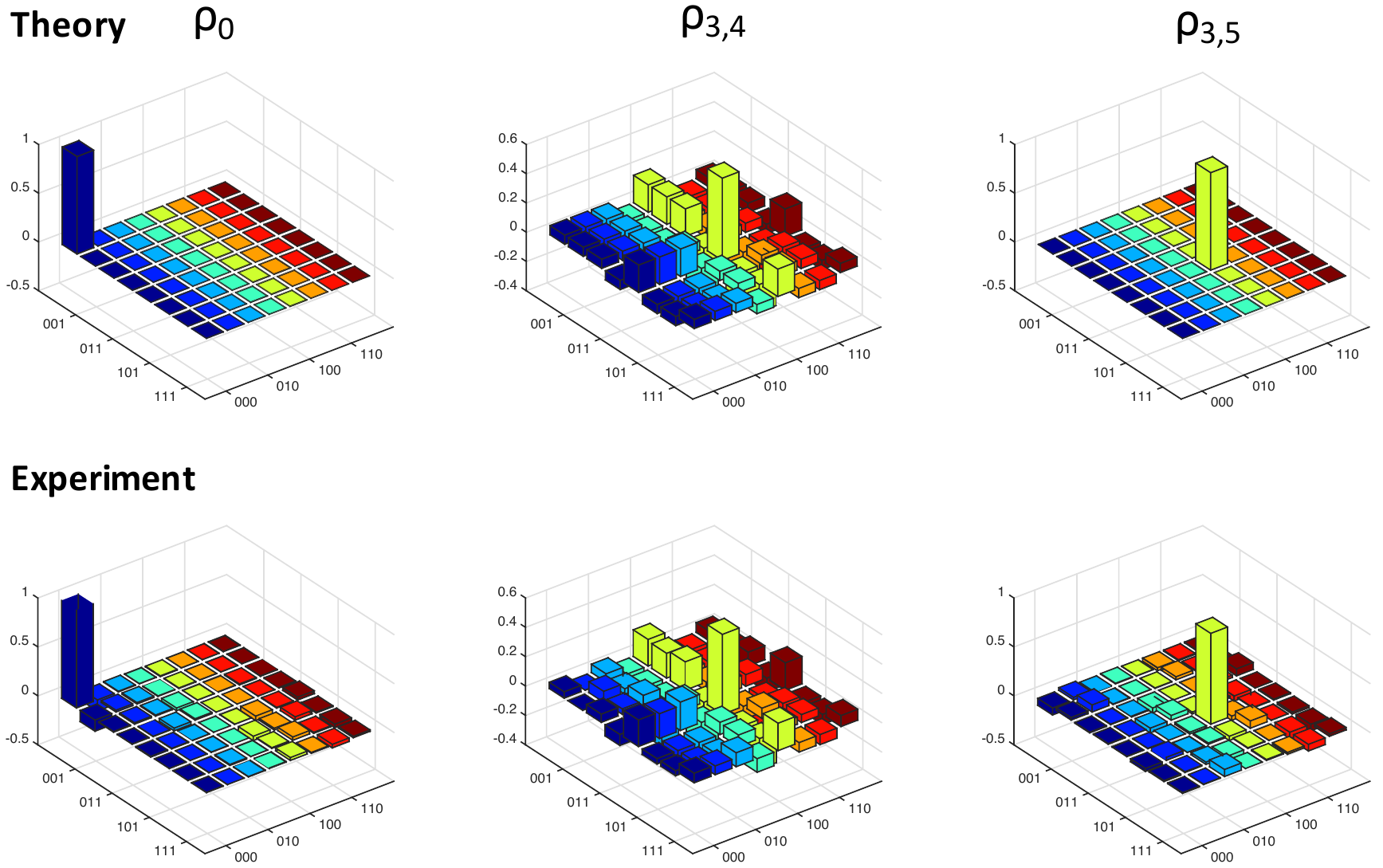}
\caption{Real parts of theoretical and experimental density matrices of $\rho_0$, $\rho_{3,4}$ and $\rho_{3,5}$. }\label{f6}
\end{center}
\end{figure}

QST is a useful technique to reconstruct and characterize the quantum state for our system. The initial state of our system (i.e., PPS) $\rho_0$ and the final state of 2 selected instance (instance 4 and 5) in 3-fold \textsc{Forrelation} case $\rho_{3,4}$ and $\rho_{3,5}$ after implementing the circuit are reconstructed and shown in Fig.~\ref{f6} (Imaginary parts are not shown).

\begin{thebibliography}{35}%
\makeatletter
\providecommand \@ifxundefined [1]{%
 \@ifx{#1\undefined}
}%
\providecommand \@ifnum [1]{%
 \ifnum #1\expandafter \@firstoftwo
 \else \expandafter \@secondoftwo
 \fi
}%
\providecommand \@ifx [1]{%
 \ifx #1\expandafter \@firstoftwo
 \else \expandafter \@secondoftwo
 \fi
}%
\providecommand \natexlab [1]{#1}%
\providecommand \enquote  [1]{``#1''}%
\providecommand \bibnamefont  [1]{#1}%
\providecommand \bibfnamefont [1]{#1}%
\providecommand \citenamefont [1]{#1}%
\providecommand \href@noop [0]{\@secondoftwo}%
\providecommand \href [0]{\begingroup \@sanitize@url \@href}%
\providecommand \@href[1]{\@@startlink{#1}\@@href}%
\providecommand \@@href[1]{\endgroup#1\@@endlink}%
\providecommand \@sanitize@url [0]{\catcode `\\12\catcode `\$12\catcode
  `\&12\catcode `\#12\catcode `\^12\catcode `\_12\catcode `\%12\relax}%
\providecommand \@@startlink[1]{}%
\providecommand \@@endlink[0]{}%
\providecommand \url  [0]{\begingroup\@sanitize@url \@url }%
\providecommand \@url [1]{\endgroup\@href {#1}{\urlprefix }}%
\providecommand \urlprefix  [0]{URL }%
\providecommand \Eprint [0]{\href }%
\providecommand \doibase [0]{http://dx.doi.org/}%
\providecommand \selectlanguage [0]{\@gobble}%
\providecommand \bibinfo  [0]{\@secondoftwo}%
\providecommand \bibfield  [0]{\@secondoftwo}%
\providecommand \translation [1]{[#1]}%
\providecommand \BibitemOpen [0]{}%
\providecommand \bibitemStop [0]{}%
\providecommand \bibitemNoStop [0]{.\EOS\space}%
\providecommand \EOS [0]{\spacefactor3000\relax}%
\providecommand \BibitemShut  [1]{\csname bibitem#1\endcsname}%
\let\auto@bib@innerbib\@empty
\bibitem [{\citenamefont {Shor}(1994)}]{shor1994algorithms}%
  \BibitemOpen
  \bibfield  {author} {\bibinfo {author} {\bibfnamefont {P.~W.}\ \bibnamefont
  {Shor}},\ }in\ \href {\doibase 10.1109/SFCS.1994.365700} {\emph {\bibinfo
  {booktitle} {Proceedings 35th Annual Symposium on Foundations of Computer
  Science}}}\ (\bibinfo {year} {1994})\ pp.\ \bibinfo {pages}
  {124--134}\BibitemShut {NoStop}%
\bibitem [{\citenamefont {Harrow}\ \emph {et~al.}(2009)\citenamefont {Harrow},
  \citenamefont {Hassidim},\ and\ \citenamefont {Lloyd}}]{harrow2009quantum}%
  \BibitemOpen
  \bibfield  {author} {\bibinfo {author} {\bibfnamefont {A.~W.}\ \bibnamefont
  {Harrow}}, \bibinfo {author} {\bibfnamefont {A.}~\bibnamefont {Hassidim}}, \
  and\ \bibinfo {author} {\bibfnamefont {S.}~\bibnamefont {Lloyd}},\ }\href
  {\doibase 10.1103/PhysRevLett.103.150502} {\bibfield  {journal} {\bibinfo
  {journal} {Phys. Rev. Lett.}\ }\textbf {\bibinfo {volume} {103}},\ \bibinfo
  {pages} {150502} (\bibinfo {year} {2009})}\BibitemShut {NoStop}%
\bibitem [{\citenamefont {Lloyd}(1996)}]{Lloyd1996}%
  \BibitemOpen
  \bibfield  {author} {\bibinfo {author} {\bibfnamefont {S.}~\bibnamefont
  {Lloyd}},\ }\href {\doibase 10.1126/science.273.5278.1073} {\bibfield
  {journal} {\bibinfo  {journal} {Science}\ }\textbf {\bibinfo {volume}
  {273}},\ \bibinfo {pages} {1073} (\bibinfo {year} {1996})}\BibitemShut
  {NoStop}%
\bibitem [{\citenamefont {Yung}\ \emph {et~al.}(2014)\citenamefont {Yung},
  \citenamefont {Whitfield}, \citenamefont {Boixo}, \citenamefont {Tempel},\
  and\ \citenamefont {Aspuru-Guzik}}]{Yung2012c}%
  \BibitemOpen
  \bibfield  {author} {\bibinfo {author} {\bibfnamefont {M.-H.}\ \bibnamefont
  {Yung}}, \bibinfo {author} {\bibfnamefont {J.~D.}\ \bibnamefont {Whitfield}},
  \bibinfo {author} {\bibfnamefont {S.}~\bibnamefont {Boixo}}, \bibinfo
  {author} {\bibfnamefont {D.~G.}\ \bibnamefont {Tempel}}, \ and\ \bibinfo
  {author} {\bibfnamefont {A.}~\bibnamefont {Aspuru-Guzik}},\ }in\ \href
  {\doibase 10.1002/9781118742631.ch03} {\emph {\bibinfo {booktitle} {Adv.
  Chem. Phys.}}},\ \bibinfo {series} {Advances in Chemical Physics}, Vol.\
  \bibinfo {volume} {154},\ \bibinfo {editor} {edited by\ \bibinfo {editor}
  {\bibfnamefont {S.}~\bibnamefont {Kais}}}\ (\bibinfo  {publisher} {John Wiley
  {\&} Sons, Inc.},\ \bibinfo {address} {Hoboken, New Jersey},\ \bibinfo {year}
  {2014})\ pp.\ \bibinfo {pages} {67--106}\BibitemShut {NoStop}%
\bibitem [{\citenamefont {Kassal}\ \emph {et~al.}(2011)\citenamefont {Kassal},
  \citenamefont {Whitfield}, \citenamefont {Perdomo-Ortiz}, \citenamefont
  {Yung},\ and\ \citenamefont {Aspuru-Guzik}}]{Kassal2011}%
  \BibitemOpen
  \bibfield  {author} {\bibinfo {author} {\bibfnamefont {I.}~\bibnamefont
  {Kassal}}, \bibinfo {author} {\bibfnamefont {J.~D.}\ \bibnamefont
  {Whitfield}}, \bibinfo {author} {\bibfnamefont {A.}~\bibnamefont
  {Perdomo-Ortiz}}, \bibinfo {author} {\bibfnamefont {M.-H.}\ \bibnamefont
  {Yung}}, \ and\ \bibinfo {author} {\bibfnamefont {A.}~\bibnamefont
  {Aspuru-Guzik}},\ }\href {\doibase 10.1146/annurev-physchem-032210-103512}
  {\bibfield  {journal} {\bibinfo  {journal} {Annu. Rev. Phys. Chem.}\ }\textbf
  {\bibinfo {volume} {62}},\ \bibinfo {pages} {185} (\bibinfo {year}
  {2011})}\BibitemShut {NoStop}%
\bibitem [{\citenamefont {Aaronson}\ and\ \citenamefont
  {Arkhipov}(2011)}]{aaronson2011computational}%
  \BibitemOpen
  \bibfield  {author} {\bibinfo {author} {\bibfnamefont {S.}~\bibnamefont
  {Aaronson}}\ and\ \bibinfo {author} {\bibfnamefont {A.}~\bibnamefont
  {Arkhipov}},\ }in\ \href {\doibase 10.1145/1993636.1993682} {\emph {\bibinfo
  {booktitle} {Proceedings of the Forty-third Annual ACM Symposium on Theory of
  Computing}}},\ \bibinfo {series and number} {STOC '11}\ (\bibinfo
  {publisher} {ACM},\ \bibinfo {address} {New York, NY, USA},\ \bibinfo {year}
  {2011})\ pp.\ \bibinfo {pages} {333--342}\BibitemShut {NoStop}%
\bibitem [{\citenamefont {Bremner}\ \emph {et~al.}(2010)\citenamefont
  {Bremner}, \citenamefont {Jozsa},\ and\ \citenamefont
  {Shepherd}}]{bremner2011classical}%
  \BibitemOpen
  \bibfield  {author} {\bibinfo {author} {\bibfnamefont {M.~J.}\ \bibnamefont
  {Bremner}}, \bibinfo {author} {\bibfnamefont {R.}~\bibnamefont {Jozsa}}, \
  and\ \bibinfo {author} {\bibfnamefont {D.~J.}\ \bibnamefont {Shepherd}},\
  }\href {\doibase 10.1098/rspa.2010.0301} {\bibfield  {journal} {\bibinfo
  {journal} {Proc. R. Soc. A}\ }\textbf {\bibinfo {volume} {467}},\ \bibinfo
  {pages} {459} (\bibinfo {year} {2010})}\BibitemShut {NoStop}%
\bibitem [{\citenamefont {Bremner}\ \emph {et~al.}()\citenamefont {Bremner},
  \citenamefont {Montanaro},\ and\ \citenamefont
  {Shepherd}}]{bremner2016achieving}%
  \BibitemOpen
  \bibfield  {author} {\bibinfo {author} {\bibfnamefont {M.~J.}\ \bibnamefont
  {Bremner}}, \bibinfo {author} {\bibfnamefont {A.}~\bibnamefont {Montanaro}},
  \ and\ \bibinfo {author} {\bibfnamefont {D.~J.}\ \bibnamefont {Shepherd}},\
  }\href {https://arxiv.org/abs/1610.01808} {\bibinfo  {journal}
  {arXiv:1610.01808}\ }\BibitemShut {NoStop}%
\bibitem [{\citenamefont {Gao}\ \emph {et~al.}()\citenamefont {Gao},
  \citenamefont {Wang},\ and\ \citenamefont {Duan}}]{gao2016quantum}%
  \BibitemOpen
\bibfield  {journal} {  }\bibfield  {author} {\bibinfo {author} {\bibfnamefont
  {X.}~\bibnamefont {Gao}}, \bibinfo {author} {\bibfnamefont {S.-T.}\
  \bibnamefont {Wang}}, \ and\ \bibinfo {author} {\bibfnamefont {L.-M.}\
  \bibnamefont {Duan}},\ }\href {https://arxiv.org/abs/1607.04947} {\bibinfo
  {journal} {arXiv:1607.04947}\ }\BibitemShut {NoStop}%
\bibitem [{\citenamefont {Boixo}\ \emph {et~al.}()\citenamefont {Boixo},
  \citenamefont {Isakov}, \citenamefont {Smelyanskiy}, \citenamefont {Babbush},
  \citenamefont {Ding}, \citenamefont {Jiang}, \citenamefont {Martinis},\ and\
  \citenamefont {Neven}}]{boixo2016characterizing}%
  \BibitemOpen
\bibfield  {journal} {  }\bibfield  {author} {\bibinfo {author} {\bibfnamefont
  {S.}~\bibnamefont {Boixo}}, \bibinfo {author} {\bibfnamefont {S.~V.}\
  \bibnamefont {Isakov}}, \bibinfo {author} {\bibfnamefont {V.~N.}\
  \bibnamefont {Smelyanskiy}}, \bibinfo {author} {\bibfnamefont
  {R.}~\bibnamefont {Babbush}}, \bibinfo {author} {\bibfnamefont
  {N.}~\bibnamefont {Ding}}, \bibinfo {author} {\bibfnamefont {Z.}~\bibnamefont
  {Jiang}}, \bibinfo {author} {\bibfnamefont {J.~M.}\ \bibnamefont {Martinis}},
  \ and\ \bibinfo {author} {\bibfnamefont {H.}~\bibnamefont {Neven}},\ }\href
  {https://arxiv.org/abs/1608.00263} {\bibinfo  {journal} {arXiv:1608.00263}\
  }\BibitemShut {NoStop}%
\bibitem [{\citenamefont {{Preskill}}()}]{preskill2012quantum}%
  \BibitemOpen
\bibfield  {journal} {  }\bibfield  {author} {\bibinfo {author} {\bibfnamefont
  {J.}~\bibnamefont {{Preskill}}},\ }\href {https://arxiv.org/abs/1203.5813}
  {\bibinfo  {journal} {arXiv:1203.5813}\ }\BibitemShut {NoStop}%
\bibitem [{\citenamefont {Yung}\ \emph {et~al.}()\citenamefont {Yung},
  \citenamefont {Gao},\ and\ \citenamefont {Huh}}]{Yung2016}%
  \BibitemOpen
\bibfield  {journal} {  }\bibfield  {author} {\bibinfo {author} {\bibfnamefont
  {M.-H.}\ \bibnamefont {Yung}}, \bibinfo {author} {\bibfnamefont
  {X.}~\bibnamefont {Gao}}, \ and\ \bibinfo {author} {\bibfnamefont
  {J.}~\bibnamefont {Huh}},\ }\href {http://arxiv.org/abs/1608.00383} {\bibinfo
   {journal} {arXiv:1608.00383}\ }\BibitemShut {NoStop}%
\bibitem [{\citenamefont {Grover}(1996)}]{grover1996fast}%
  \BibitemOpen
\bibfield  {journal} {  }\bibfield  {author} {\bibinfo {author} {\bibfnamefont
  {L.~K.}\ \bibnamefont {Grover}},\ }in\ \href {\doibase 10.1145/237814.237866}
  {\emph {\bibinfo {booktitle} {Proceedings of the Twenty-eighth Annual ACM
  Symposium on Theory of Computing}}},\ \bibinfo {series and number} {STOC
  '96}\ (\bibinfo  {publisher} {ACM},\ \bibinfo {address} {New York, NY, USA},\
  \bibinfo {year} {1996})\ pp.\ \bibinfo {pages} {212--219}\BibitemShut
  {NoStop}%
\bibitem [{\citenamefont {Deutsch}\ and\ \citenamefont
  {Jozsa}(1992)}]{deutsch1992rapid}%
  \BibitemOpen
  \bibfield  {author} {\bibinfo {author} {\bibfnamefont {D.}~\bibnamefont
  {Deutsch}}\ and\ \bibinfo {author} {\bibfnamefont {R.}~\bibnamefont
  {Jozsa}},\ }\href {\doibase 10.1098/rspa.1992.0167} {\bibfield  {journal}
  {\bibinfo  {journal} {Proceedings of the Royal Society of London A:
  Mathematical, Physical and Engineering Sciences}\ }\textbf {\bibinfo {volume}
  {439}},\ \bibinfo {pages} {553} (\bibinfo {year} {1992})}\BibitemShut
  {NoStop}%
\bibitem [{\citenamefont {Simon}(1997)}]{simon1997power}%
  \BibitemOpen
  \bibfield  {author} {\bibinfo {author} {\bibfnamefont {D.~R.}\ \bibnamefont
  {Simon}},\ }\href {\doibase 10.1137/S0097539796298637} {\bibfield  {journal}
  {\bibinfo  {journal} {SIAM Journal on Computing}\ }\textbf {\bibinfo {volume}
  {26}},\ \bibinfo {pages} {1474} (\bibinfo {year} {1997})}\BibitemShut
  {NoStop}%
\bibitem [{\citenamefont {Aaronson}\ and\ \citenamefont
  {Ambainis}(2015)}]{aaronson2015forrelation}%
  \BibitemOpen
  \bibfield  {author} {\bibinfo {author} {\bibfnamefont {S.}~\bibnamefont
  {Aaronson}}\ and\ \bibinfo {author} {\bibfnamefont {A.}~\bibnamefont
  {Ambainis}},\ }in\ \href {\doibase 10.1145/2746539.2746547} {\emph {\bibinfo
  {booktitle} {Proceedings of the Forty-seventh Annual ACM Symposium on Theory
  of Computing}}},\ \bibinfo {series and number} {STOC '15}\ (\bibinfo
  {publisher} {ACM},\ \bibinfo {address} {New York, NY, USA},\ \bibinfo {year}
  {2015})\ pp.\ \bibinfo {pages} {307--316}\BibitemShut {NoStop}%
\bibitem [{\citenamefont {Aaronson}(2010)}]{aaronson2010bqp}%
  \BibitemOpen
  \bibfield  {author} {\bibinfo {author} {\bibfnamefont {S.}~\bibnamefont
  {Aaronson}},\ }in\ \href {\doibase 10.1145/1806689.1806711} {\emph {\bibinfo
  {booktitle} {Proceedings of the Forty-second ACM Symposium on Theory of
  Computing}}},\ \bibinfo {series and number} {STOC '10}\ (\bibinfo
  {publisher} {ACM},\ \bibinfo {address} {New York, NY, USA},\ \bibinfo {year}
  {2010})\ pp.\ \bibinfo {pages} {141--150}\BibitemShut {NoStop}%
\bibitem [{\citenamefont {De~Beaudrap}\ \emph {et~al.}(2002)\citenamefont
  {De~Beaudrap}, \citenamefont {Cleve},\ and\ \citenamefont
  {Watrous}}]{de2002sharp}%
  \BibitemOpen
  \bibfield  {author} {\bibinfo {author} {\bibfnamefont {J.~N.}\ \bibnamefont
  {De~Beaudrap}}, \bibinfo {author} {\bibfnamefont {R.}~\bibnamefont {Cleve}},
  \ and\ \bibinfo {author} {\bibfnamefont {J.}~\bibnamefont {Watrous}},\ }\href
  {\doibase 10.1007/s00453-002-0978-1} {\bibfield  {journal} {\bibinfo
  {journal} {Algorithmica}\ }\textbf {\bibinfo {volume} {34}},\ \bibinfo
  {pages} {449} (\bibinfo {year} {2002})}\BibitemShut {NoStop}%
\bibitem [{\citenamefont {Chakraborty}\ \emph {et~al.}()\citenamefont
  {Chakraborty}, \citenamefont {Fischer}, \citenamefont {Matsliah},\ and\
  \citenamefont {De~Wolf}}]{chakraborty2010new}%
  \BibitemOpen
  \bibfield  {author} {\bibinfo {author} {\bibfnamefont {S.}~\bibnamefont
  {Chakraborty}}, \bibinfo {author} {\bibfnamefont {E.}~\bibnamefont
  {Fischer}}, \bibinfo {author} {\bibfnamefont {A.}~\bibnamefont {Matsliah}}, \
  and\ \bibinfo {author} {\bibfnamefont {R.}~\bibnamefont {De~Wolf}},\ }\href
  {https://arxiv.org/abs/1005.0523} {\bibinfo  {journal} {arXiv:1005.0523}\
  }\BibitemShut {NoStop}%
\bibitem [{\citenamefont {Chuang}\ \emph {et~al.}(1998)\citenamefont {Chuang},
  \citenamefont {Vandersypen}, \citenamefont {Zhou}, \citenamefont {Leung},\
  and\ \citenamefont {Lloyd}}]{chuang1998experimental}%
  \BibitemOpen
\bibfield  {journal} {  }\bibfield  {author} {\bibinfo {author} {\bibfnamefont
  {I.~L.}\ \bibnamefont {Chuang}}, \bibinfo {author} {\bibfnamefont {L.~M.}\
  \bibnamefont {Vandersypen}}, \bibinfo {author} {\bibfnamefont
  {X.}~\bibnamefont {Zhou}}, \bibinfo {author} {\bibfnamefont {D.~W.}\
  \bibnamefont {Leung}}, \ and\ \bibinfo {author} {\bibfnamefont
  {S.}~\bibnamefont {Lloyd}},\ }\href {http://dx.doi.org/10.1038/30181}
  {\bibfield  {journal} {\bibinfo  {journal} {Nature}\ }\textbf {\bibinfo
  {volume} {393}},\ \bibinfo {pages} {143} (\bibinfo {year}
  {1998})}\BibitemShut {NoStop}%
\bibitem [{\citenamefont {Jones}\ and\ \citenamefont
  {Mosca}(1998)}]{jones1998implementation}%
  \BibitemOpen
  \bibfield  {author} {\bibinfo {author} {\bibfnamefont {J.~A.}\ \bibnamefont
  {Jones}}\ and\ \bibinfo {author} {\bibfnamefont {M.}~\bibnamefont {Mosca}},\
  }\href {http://dx.doi.org/10.1063/1.476739} {\bibfield  {journal} {\bibinfo
  {journal} {J. Chem. Phys.}\ }\textbf {\bibinfo {volume} {109}},\ \bibinfo
  {pages} {1648} (\bibinfo {year} {1998})}\BibitemShut {NoStop}%
\bibitem [{\citenamefont {Peng}\ \emph {et~al.}(2015)\citenamefont {Peng},
  \citenamefont {Zhou}, \citenamefont {Wei}, \citenamefont {Cui}, \citenamefont
  {Du},\ and\ \citenamefont {Liu}}]{peng2015zeros}%
  \BibitemOpen
  \bibfield  {author} {\bibinfo {author} {\bibfnamefont {X.}~\bibnamefont
  {Peng}}, \bibinfo {author} {\bibfnamefont {H.}~\bibnamefont {Zhou}}, \bibinfo
  {author} {\bibfnamefont {B.-B.}\ \bibnamefont {Wei}}, \bibinfo {author}
  {\bibfnamefont {J.}~\bibnamefont {Cui}}, \bibinfo {author} {\bibfnamefont
  {J.}~\bibnamefont {Du}}, \ and\ \bibinfo {author} {\bibfnamefont {R.-B.}\
  \bibnamefont {Liu}},\ }\href {\doibase 10.1103/PhysRevLett.114.010601}
  {\bibfield  {journal} {\bibinfo  {journal} {Phys. Rev. Lett.}\ }\textbf
  {\bibinfo {volume} {114}},\ \bibinfo {pages} {010601} (\bibinfo {year}
  {2015})}\BibitemShut {NoStop}%
\bibitem [{\citenamefont {Lu}\ \emph {et~al.}(2015)\citenamefont {Lu},
  \citenamefont {Li}, \citenamefont {Trottier}, \citenamefont {Li},
  \citenamefont {Brodutch}, \citenamefont {Krismanich}, \citenamefont
  {Ghavami}, \citenamefont {Dmitrienko}, \citenamefont {Long}, \citenamefont
  {Baugh},\ and\ \citenamefont {Laflamme}}]{lu2015clifford}%
  \BibitemOpen
  \bibfield  {author} {\bibinfo {author} {\bibfnamefont {D.}~\bibnamefont
  {Lu}}, \bibinfo {author} {\bibfnamefont {H.}~\bibnamefont {Li}}, \bibinfo
  {author} {\bibfnamefont {D.-A.}\ \bibnamefont {Trottier}}, \bibinfo {author}
  {\bibfnamefont {J.}~\bibnamefont {Li}}, \bibinfo {author} {\bibfnamefont
  {A.}~\bibnamefont {Brodutch}}, \bibinfo {author} {\bibfnamefont {A.~P.}\
  \bibnamefont {Krismanich}}, \bibinfo {author} {\bibfnamefont
  {A.}~\bibnamefont {Ghavami}}, \bibinfo {author} {\bibfnamefont {G.~I.}\
  \bibnamefont {Dmitrienko}}, \bibinfo {author} {\bibfnamefont
  {G.}~\bibnamefont {Long}}, \bibinfo {author} {\bibfnamefont {J.}~\bibnamefont
  {Baugh}}, \ and\ \bibinfo {author} {\bibfnamefont {R.}~\bibnamefont
  {Laflamme}},\ }\href {\doibase 10.1103/PhysRevLett.114.140505} {\bibfield
  {journal} {\bibinfo  {journal} {Phys. Rev. Lett.}\ }\textbf {\bibinfo
  {volume} {114}},\ \bibinfo {pages} {140505} (\bibinfo {year}
  {2015})}\BibitemShut {NoStop}%
\bibitem [{\citenamefont {Feng}\ \emph {et~al.}(2013)\citenamefont {Feng},
  \citenamefont {Lu}, \citenamefont {Hao}, \citenamefont {Zhang},\ and\
  \citenamefont {Long}}]{feng2013tunnel}%
  \BibitemOpen
  \bibfield  {author} {\bibinfo {author} {\bibfnamefont {G.-R.}\ \bibnamefont
  {Feng}}, \bibinfo {author} {\bibfnamefont {Y.}~\bibnamefont {Lu}}, \bibinfo
  {author} {\bibfnamefont {L.}~\bibnamefont {Hao}}, \bibinfo {author}
  {\bibfnamefont {F.-H.}\ \bibnamefont {Zhang}}, \ and\ \bibinfo {author}
  {\bibfnamefont {G.-L.}\ \bibnamefont {Long}},\ }\href
  {http://www.nature.com/articles/srep02232} {\bibfield  {journal} {\bibinfo
  {journal} {Sci. Rep.}\ }\textbf {\bibinfo {volume} {3}},\ \bibinfo {pages}
  {2232} (\bibinfo {year} {2013})}\BibitemShut {NoStop}%
\bibitem [{\citenamefont {Peng}\ \emph {et~al.}(2005)\citenamefont {Peng},
  \citenamefont {Du},\ and\ \citenamefont {Suter}}]{peng2005phase}%
  \BibitemOpen
  \bibfield  {author} {\bibinfo {author} {\bibfnamefont {X.}~\bibnamefont
  {Peng}}, \bibinfo {author} {\bibfnamefont {J.}~\bibnamefont {Du}}, \ and\
  \bibinfo {author} {\bibfnamefont {D.}~\bibnamefont {Suter}},\ }\href
  {\doibase 10.1103/PhysRevA.71.012307} {\bibfield  {journal} {\bibinfo
  {journal} {Phys. Rev. A}\ }\textbf {\bibinfo {volume} {71}},\ \bibinfo
  {pages} {012307} (\bibinfo {year} {2005})}\BibitemShut {NoStop}%
\bibitem [{\citenamefont {Auccaise}\ \emph {et~al.}(2011)\citenamefont
  {Auccaise}, \citenamefont {C\'eleri}, \citenamefont {Soares-Pinto},
  \citenamefont {deAzevedo}, \citenamefont {Maziero}, \citenamefont {Souza},
  \citenamefont {Bonagamba}, \citenamefont {Sarthour}, \citenamefont
  {Oliveira},\ and\ \citenamefont {Serra}}]{auccaise2011discord}%
  \BibitemOpen
  \bibfield  {author} {\bibinfo {author} {\bibfnamefont {R.}~\bibnamefont
  {Auccaise}}, \bibinfo {author} {\bibfnamefont {L.~C.}\ \bibnamefont
  {C\'eleri}}, \bibinfo {author} {\bibfnamefont {D.~O.}\ \bibnamefont
  {Soares-Pinto}}, \bibinfo {author} {\bibfnamefont {E.~R.}\ \bibnamefont
  {deAzevedo}}, \bibinfo {author} {\bibfnamefont {J.}~\bibnamefont {Maziero}},
  \bibinfo {author} {\bibfnamefont {A.~M.}\ \bibnamefont {Souza}}, \bibinfo
  {author} {\bibfnamefont {T.~J.}\ \bibnamefont {Bonagamba}}, \bibinfo {author}
  {\bibfnamefont {R.~S.}\ \bibnamefont {Sarthour}}, \bibinfo {author}
  {\bibfnamefont {I.~S.}\ \bibnamefont {Oliveira}}, \ and\ \bibinfo {author}
  {\bibfnamefont {R.~M.}\ \bibnamefont {Serra}},\ }\href {\doibase
  10.1103/PhysRevLett.107.140403} {\bibfield  {journal} {\bibinfo  {journal}
  {Phys. Rev. Lett.}\ }\textbf {\bibinfo {volume} {107}},\ \bibinfo {pages}
  {140403} (\bibinfo {year} {2011})}\BibitemShut {NoStop}%
\bibitem [{\citenamefont {Zhang}\ \emph {et~al.}(2012)\citenamefont {Zhang},
  \citenamefont {Yung}, \citenamefont {Laflamme}, \citenamefont
  {Aspuru-Guzik},\ and\ \citenamefont {Baugh}}]{zhang2012magnet}%
  \BibitemOpen
  \bibfield  {author} {\bibinfo {author} {\bibfnamefont {J.}~\bibnamefont
  {Zhang}}, \bibinfo {author} {\bibfnamefont {M.-H.}\ \bibnamefont {Yung}},
  \bibinfo {author} {\bibfnamefont {R.}~\bibnamefont {Laflamme}}, \bibinfo
  {author} {\bibfnamefont {A.}~\bibnamefont {Aspuru-Guzik}}, \ and\ \bibinfo
  {author} {\bibfnamefont {J.}~\bibnamefont {Baugh}},\ }\href
  {http://dx.doi.org/10.1038/ncomms1860} {\bibfield  {journal} {\bibinfo
  {journal} {Nat. Commun.}\ }\textbf {\bibinfo {volume} {3}},\ \bibinfo {pages}
  {880} (\bibinfo {year} {2012})}\BibitemShut {NoStop}%
\bibitem [{\citenamefont {Gershenfeld}\ and\ \citenamefont
  {Chuang}(1997)}]{gershenfeld1997qc}%
  \BibitemOpen
  \bibfield  {author} {\bibinfo {author} {\bibfnamefont {N.~A.}\ \bibnamefont
  {Gershenfeld}}\ and\ \bibinfo {author} {\bibfnamefont {I.~L.}\ \bibnamefont
  {Chuang}},\ }\href {\doibase 10.1126/science.275.5298.350} {\bibfield
  {journal} {\bibinfo  {journal} {Science}\ }\textbf {\bibinfo {volume}
  {275}},\ \bibinfo {pages} {350} (\bibinfo {year} {1997})}\BibitemShut
  {NoStop}%
\bibitem [{\citenamefont {Knill}\ \emph {et~al.}(2000)\citenamefont {Knill},
  \citenamefont {Laflamme}, \citenamefont {Martinez},\ and\ \citenamefont
  {Tseng}}]{knill2000algorithmic}%
  \BibitemOpen
  \bibfield  {author} {\bibinfo {author} {\bibfnamefont {E.}~\bibnamefont
  {Knill}}, \bibinfo {author} {\bibfnamefont {R.}~\bibnamefont {Laflamme}},
  \bibinfo {author} {\bibfnamefont {R.}~\bibnamefont {Martinez}}, \ and\
  \bibinfo {author} {\bibfnamefont {C.-H.}\ \bibnamefont {Tseng}},\ }\href
  {http://dx.doi.org/10.1038/35006012} {\bibfield  {journal} {\bibinfo
  {journal} {Nature}\ }\textbf {\bibinfo {volume} {404}},\ \bibinfo {pages}
  {368} (\bibinfo {year} {2000})}\BibitemShut {NoStop}%
\bibitem [{\citenamefont {Cory}\ \emph {et~al.}(1997)\citenamefont {Cory},
  \citenamefont {Fahmy},\ and\ \citenamefont {Havel}}]{cory1997spatial}%
  \BibitemOpen
  \bibfield  {author} {\bibinfo {author} {\bibfnamefont {D.~G.}\ \bibnamefont
  {Cory}}, \bibinfo {author} {\bibfnamefont {A.~F.}\ \bibnamefont {Fahmy}}, \
  and\ \bibinfo {author} {\bibfnamefont {T.~F.}\ \bibnamefont {Havel}},\ }\href
  {http://www.pnas.org/content/94/5/1634.abstract} {\bibfield  {journal}
  {\bibinfo  {journal} {Proc. Natl. Acad. Sci.}\ }\textbf {\bibinfo {volume}
  {94}},\ \bibinfo {pages} {1634} (\bibinfo {year} {1997})}\BibitemShut
  {NoStop}%
\bibitem [{\citenamefont {Lee}(2002)}]{lee2002nmr}%
  \BibitemOpen
  \bibfield  {author} {\bibinfo {author} {\bibfnamefont {J.-S.}\ \bibnamefont
  {Lee}},\ }\href {\doibase http://dx.doi.org/10.1016/S0375-9601(02)01479-2}
  {\bibfield  {journal} {\bibinfo  {journal} {Phys. Lett. A}\ }\textbf
  {\bibinfo {volume} {305}},\ \bibinfo {pages} {349 } (\bibinfo {year}
  {2002})}\BibitemShut {NoStop}%
\bibitem [{\citenamefont {Vandersypen}\ and\ \citenamefont
  {Chuang}(2005)}]{vandersypen2005nmr}%
  \BibitemOpen
  \bibfield  {author} {\bibinfo {author} {\bibfnamefont {L.~M.~K.}\
  \bibnamefont {Vandersypen}}\ and\ \bibinfo {author} {\bibfnamefont {I.~L.}\
  \bibnamefont {Chuang}},\ }\href {\doibase 10.1103/RevModPhys.76.1037}
  {\bibfield  {journal} {\bibinfo  {journal} {Rev. Mod. Phys.}\ }\textbf
  {\bibinfo {volume} {76}},\ \bibinfo {pages} {1037} (\bibinfo {year}
  {2005})}\BibitemShut {NoStop}%
\bibitem [{\citenamefont {Ryan}\ \emph {et~al.}(2008)\citenamefont {Ryan},
  \citenamefont {Negrevergne}, \citenamefont {Laforest}, \citenamefont
  {Knill},\ and\ \citenamefont {Laflamme}}]{ryan2008compiler}%
  \BibitemOpen
  \bibfield  {author} {\bibinfo {author} {\bibfnamefont {C.~A.}\ \bibnamefont
  {Ryan}}, \bibinfo {author} {\bibfnamefont {C.}~\bibnamefont {Negrevergne}},
  \bibinfo {author} {\bibfnamefont {M.}~\bibnamefont {Laforest}}, \bibinfo
  {author} {\bibfnamefont {E.}~\bibnamefont {Knill}}, \ and\ \bibinfo {author}
  {\bibfnamefont {R.}~\bibnamefont {Laflamme}},\ }\href {\doibase
  10.1103/PhysRevA.78.012328} {\bibfield  {journal} {\bibinfo  {journal} {Phys.
  Rev. A}\ }\textbf {\bibinfo {volume} {78}},\ \bibinfo {pages} {012328}
  (\bibinfo {year} {2008})}\BibitemShut {NoStop}%
\bibitem [{\citenamefont {Nielsen}\ and\ \citenamefont
  {Chuang}(2010)}]{nielsen2010quantum}%
  \BibitemOpen
  \bibfield  {author} {\bibinfo {author} {\bibfnamefont {M.~A.}\ \bibnamefont
  {Nielsen}}\ and\ \bibinfo {author} {\bibfnamefont {I.~L.}\ \bibnamefont
  {Chuang}},\ }\href@noop {} {\emph {\bibinfo {title} {Quantum computation and
  quantum information}}}\ (\bibinfo  {publisher} {Cambridge university press},\
  \bibinfo {year} {2010})\BibitemShut {NoStop}%
\bibitem [{\citenamefont {Khaneja}\ \emph {et~al.}(2005)\citenamefont
  {Khaneja}, \citenamefont {Reiss}, \citenamefont {Kehlet}, \citenamefont
  {Schulte-Herbrüggen},\ and\ \citenamefont {Glaser}}]{khaneja2005grape}%
  \BibitemOpen
  \bibfield  {author} {\bibinfo {author} {\bibfnamefont {N.}~\bibnamefont
  {Khaneja}}, \bibinfo {author} {\bibfnamefont {T.}~\bibnamefont {Reiss}},
  \bibinfo {author} {\bibfnamefont {C.}~\bibnamefont {Kehlet}}, \bibinfo
  {author} {\bibfnamefont {T.}~\bibnamefont {Schulte-Herbrüggen}}, \ and\
  \bibinfo {author} {\bibfnamefont {S.~J.}\ \bibnamefont {Glaser}},\ }\href
  {\doibase http://dx.doi.org/10.1016/j.jmr.2004.11.004} {\bibfield  {journal}
  {\bibinfo  {journal} {J. Magn. Reson.}\ }\textbf {\bibinfo {volume} {172}},\
  \bibinfo {pages} {296 } (\bibinfo {year} {2005})}\BibitemShut {NoStop}%
\end{thebibliography}
\end{document}